# Spin-to-charge conversion at KTaO$_3$(111) interfaces

Athby H. Al-Tawhid[1,†], Rui Sun[2,†], Andrew H. Comstock[2], Divine P. Kumah[3], Dali Sun[2], and Kaveh Ahadi[1,2,4,5*]

[1]Department of Materials Science and Engineering, North Carolina State University, Raleigh, NC 27695, USA
[2]Department of Physics, North Carolina State University, Raleigh, NC 27695, USA
[3]Department of Physics, Duke University, Durham, NC 27701, USA
[4]Department of Electrical and Computer Engineering, The Ohio State University, Columbus, OH 43210, USA
[5]Department of Materials Science and Engineering, The Ohio State University, Columbus, OH 43210, USA

[†] These authors contributed equally to this work

[*] Corresponding author. Email: ahadi.4@osu.edu

**ABSTRACT**

Rashba spin-orbit coupling locks the spin with momentum of charge carriers at the broken inversion interfaces, which could generate a large spin galvanic response. Here, we demonstrate spin-to-charge conversion (inverse Rashba-Edelstein effect) in $KTaO_3$(111) two-dimensional electron systems. We explain the results in the context of electronic structure, orbital character, and spin texture at the $KTaO_3$(111) interfaces. We also show that the angle dependence of the spin-to-charge conversion on in-plane magnetic field exhibits a nontrivial behavior which matches the symmetry of the Fermi states. Results point to opportunities to use spin-to-charge conversion as a tool to investigate the electronic structure and spin texture.

Spin-orbit coupling (SOC) enables a plethora of exciting phenomena, including charge-to-spin interconversion, with coupling the spin degree of freedom to orbital motion. Efficient charge-to-spin interconversion could potentially enable low energy, non-volatile spintronic device technologies[1]. Two-dimensional electron systems frequently exhibit superior charge-to-spin interconversion compared to bulk samples[2]. For example, 2D electron gas at the SrTiO$_3$(001) interfaces demonstrates highly efficient spin-to-charge conversion ($\lambda_{IREE}$=6.4 nm)[3]. At such interfaces combination of spin orbit coupling (SOC) and out-of-plane broken inversion symmetry elements (i.e., Rashba SOC) gives rise to a helically ordered in-plane spin configuration at which the momentum and spin degrees of freedom are locked[4]. The corresponding Rashba Hamiltonian is expressed as, $\widehat{\mathcal{H}_R} = (\alpha_R/\hbar)(\mathbf{z} \times \mathbf{p}) \cdot \boldsymbol{\sigma}$, where $\boldsymbol{\sigma}$ is the vector of the Pauli spin matrices, $\mathbf{p}$ the momentum and $\alpha_R$ the Rashba parameter. In real systems, however, interfacial spin texture frequently deviates from the ideal Rashba configuration. For example, the symmetry-restricted spin textures and orbital arrangements give rise to deviations from ideal Rashba spin-momentum locking configuration[5,6]. Furthermore, entangled spin and orbital degrees of freedom may lead to single particle states with vanishing magnetic moments[7,8].

KTaO$_3$ is a cubic perovskite and an incipient ferroelectric[9–11]. Itinerant electrons occupy the tantalum 5$d$-derived $t_{2g}$ states in electron-doped sample[12]. Tantalum 5$d$-driven conduction electrons in KTaO$_3$ have one order of magnitude larger spin–orbit coupling compared to titanium 3$d$-driven conduction electrons in EuTiO$_3$[13,14] and SrTiO$_3$[6,15], which may point to possibility of efficient charge-to-spin interconversion in KTaO$_3$ 2DEGs. Spin-charge interconversion has been reported at KTaO$_3$(001) interfaces recently[16]. The KTaO$_3$(111) interfaces, however, have not been explored, despite recent reports of exotic properties and functionalities[17,18].

Figure 1(a) exhibits the Fermi surface and spin texture of the 2DEG at the KTaO$_3$(111) interface for chemical potentials µ = 50 meV (bottom) and µ = 100 meV (top). The calculation details were described elsewhere[7]. The spin texture at KTaO$_3$(111) interfaces exhibits an anisotropic texture. The max modulo of spin (<s>) for both chemical potentials in Figure 1(a) is ~0.5 and ~0.2 for the outer band and the inner band, respectively. Previous calculations also suggest an out-of-plane canting of the otherwise helically polarized spin texture of Fermi states[6,18].

Two-dimensional electron gas (2DEG) at the interface of oxides gives rise to emergent phenomena and fundamental properties and functionalities[19–22]. A 2DEG is formed at the KTaO$_3$ (111) surface with selective removal of oxygen atoms. We deposited of ~1 nm of elemental Al from a resistive heating source. The Al tends to oxidize and creates an oxygen vacancy-induced 2DEG at the interface of the AlOx and KTaO$_3$ (111). Formation of 2DEG with deposition of a metal was initially reported by Rödel et al., depositing Al metal on SrTiO$_3$ (001) surface[23]. Oxygen vacancy-induced 2DEGs have been reported at the interface of KTaO$_3$ with AlOx[16], TiOx[24,25], EuO[17], LaCrO$_3$[26], LaAlO$_3$[27], and LaSrMnO$_3$[18]. A 100 nm amorphous SiOx was deposited immediately after the growth of Al layer to protect the 2DEG from oxidation. Thickness of each layer was calibrated using a quartz crystal microbalance (QCM), with an accuracy of ~15%, prior to the growth. NiFe (20 nm) was deposited by e-beam evaporation, instead of SiOx, immediately after the growth of Al for samples used for inverse spin Hall measurements. The growth rate for Al and NiFe were 0.1 Å/s and 0.25 Å/s, respectively.

Transport measurements were carried out using the van der Pauw geometry on square-shaped samples (5mm×5mm). Gold contacts (300 nm) were deposited on the corners of the samples using a sputtering system and a shadow mask. The magnetoelectric measurements were performed using a Quantum Design Physical Property Measurement System (PPMS). The Hall carrier density was determined from the Hall experiments, $n_{2D}=-1/(eR_H)$, where $R_H$ represents the Hall coefficient and $e$ is the elementary charge. The Hall coefficient was obtained by fitting a linear function to the transverse resistance data as a function of the magnetic field ($R_H = dR_{xy}/dB$). FMR was carried out at various temperatures with microwave frequency range of 2-16 GHz. The microwave field was generated using a commercial NanOsc PhaseFMR spectrometer with a coplanar waveguide in a PPMS. For the spin-pumping measurement, a 9-GHz microwave was produced by a Keysight X-series Microwave Analog Signal Generator with a power modulation. The transverse/longitudinal voltage was measured using a Stanford SR 830 lock in amplifier. The magnetic field and in-plane angle rotation were realized using a 6-1-1 T vector superconducting magnet.

Figure 1(b) shows the temperature dependence of the sheet resistance for SiOx/AlOx/KTaO$_3$(111) heterostructure (300-3K). The sample exhibits metallic behavior (d$R$/dT>0) and an upturn at ~12K. The room temperature Hall carrier density is 3.2×10$^{13}$ cm$^{-2}$, Figure 1(c). The 2DEG shows a carrier density freeze out below 100 K, reaching ~2.8×10$^{13}$ cm$^{-2}$ at liquid helium temperature. Figure 1(d) shows the mobility and scattering time of charge carriers with temperature exhibiting, ~60 cm$^2$/Vs and ~0.02 ps at 3 K. The observed carrier density and mobility is consistent with reported values for KTaO$_3$ (111) 2DEGs. The KTaO$_3$ (001) 2DEGs typically exhibit higher mobilities to due smoother surfaces.

To study the spin-to-charge conversion efficiency from the interfacial 2DEG, the NiFe/AlOx/ KTaO$_3$(111) sample was characterized by the broadband ferromagnetic resonance (FMR) at different temperatures. Here, the composition of NiFe is $Ni_{81}Fe_{19}$(Permalloy). Similar heterostructure has been deposited on SrTiO$_3$ (001), resolving enhanced conversion efficiencies[28]. Figure 2(a) shows the FMR spectrum at various frequencies of excitation. The resonance field and linewidth varied with microwave frequency, which can be counted by the following equation,

$$\frac{dP}{dH} = a_1 \cdot \frac{\Delta H \cdot (H-H_{res})}{4 \cdot ((H-H_{res})^2+\Delta H^2)^2} - a_2 \cdot \frac{\Delta H^2 - 4(H-H_{res})^2}{16 \cdot ((H-H_{res})^2+\Delta H^2)^2} \qquad (1)$$

where H is applied field, $H_{res}$ is the magnetic resonance field, and $\Delta H$ is full width at half maximum (FWHM) of FMR. $a_1$ and $a_2$ are the coefficient of Lorentz and anti-Lorentz line, respectively. Although, the FMR spectra exhibits strong anti-Lorentz lineshape, the fitting curves are quite well according to equation (1), from which, the resonance field and linewidth can be obtained. Figure 2(b) exhibits the resonance field with frequency, which can be described by the Kittel equation,

$$f = \frac{\mu_0 g \mu_B}{\hbar} \sqrt{(H_u + H_{res})(H_u + H_{res} + M_{eff})} \qquad (2)$$

where g is fitted to be 2.11[29] and $\mu_0 M_{eff}$ is the effective saturated magnetization with values of 0.88 T to 0.9 T as temperature decreases. The in-plane uniaxial anisotropy field $H_u$ obtained from Kittel equation exhibits a negligible value of ~1 mT at room temperature and slowly increases to 11 mT at 10 K, which could suggest presence of extended defects. Figure 2(c) shows the frequency dependence of resonance linewidth $\mu_0 \Delta H$ at 10 K, 50 K and 100 K, using the linear fitting equation below,

$$\mu_0 \Delta H = \frac{2\pi\alpha}{\gamma} f + \mu_0 \Delta H_0 \qquad (3)$$

where $\alpha$ is Gilbert damping factor which is related to the spin pumping effect from 2DEG and $\mu_0 \Delta H_0$ is due to the inhomogeneous broadening term resulting from the interfacial scattering. Gilbert damping factor $\alpha$ is just 0.0062 at 10 K, which is smaller than the resolved value at 50 K and 100 K. This could be due to the suppression of spin-phonon scattering at low temperature. The inhomogeneous broadening (and total FMR linewidth) increases below 100 K (Figure 2C). The magnetic inhomogeneities could be due to the local variations of the magnetization and anisotropy constants, potentially related to microstructural features of the samples (e.g., crystalline point and line-defects, impurities, and non-uniformity of the sample surface).

Figure 3(a) shows the power dependence of spin pumping response as a function of magnetic field at 10 K. The spin pumping response can be decomposed into symmetric part $V_{sym.}$ and antisymmetric part $V_{anti\ sym.}$, which can be expressed as

$$V_{sp} = V_{sym.} \frac{\Delta H^2}{(H-H_{res})^2 + \Delta H^2} + V_{anti\ sym.} \frac{-2\Delta H(H-H_{res})}{(H-H_{res})^2 + \Delta H^2}. \qquad (4)$$

The unintentional thermal gradient induced Seebeck signal into $V_{sym.}$ can be excluded by calculating the $V_{IEE} = [V_{sym.}(+H) - V_{sym.}(-H)]/2$. According to the standard spin pumping analysis, the spin current density $J_s^{3D}$ can be calculated through following equations[30–32].

$$g_{eff}^{\uparrow\downarrow} = \frac{\mu_0 M_{eff} t_F}{g \mu_B} (\alpha_{F/N} - \alpha_F) \qquad (5)$$

$$J_s^{3D} = \frac{g_{\text{eff}}^{\uparrow\downarrow}\gamma^2\hbar(\mu_0 h_{\text{rf}})^2}{8\pi\alpha_{eff}^2}\left[\frac{\mu_0 M_{\text{eff}}\gamma + \sqrt{(\mu_0 M_{\text{eff}}\gamma)^2 + 4\omega^2}}{(\mu_0 M_{\text{eff}}\gamma)^2 + 4\omega^2}\right]\left(\frac{2e}{\hbar}\right) \quad (6)$$

where $g_{eff}^{\uparrow\downarrow}$ is effective spin mixing conductance, $t_F$ is thickness of NiFe layer ($t_F = 20 nm$), $\alpha_{\text{NiFe}} = 0.006$ is the damping factor of reference sample 20nm-NiFe, $\mu_0 h_{\text{rf}}$ ($\mu_0 h_{\text{rf}} = 0.1 mT$) is the strength of microwave frequency field generated via the coplanar waveguides, and $\omega = 2\pi f$ (the measured frequency $f$ = 9 GHz). Figure 3(b) exhibits the spin-to-charge conversion efficiency $\lambda_{IEE}$ obtained using

$$\lambda_{IREE} = \frac{J_c^{2D}}{J_s^{3D}} = \frac{V_{IEE}}{wRJ_s^{3D}} \quad (7)$$

where w and R is the width and resistance of NiFe/AlOx/KTaO$_3$ tri-layer, and $J_c^{2D}$ is the converted charge density. The estimated $\lambda_{IEE}$ is about 0.6 nm which is larger than the traditional 3D topological insulator Bi$_2$Se$_3$[33] but remains less efficient than SrTiO$_3$ 2DEGs[3]. Here, the uniformity of $h_{RF}$, illustrated schematically in Figure 4(a), is critical for validity of Eq. 6 and, as a result, calculation of conversion efficiency. Comparison between different materials systems with similar geometry, measured using the same setup might yield more accurate quantitative comparison. Spin rectification effect (SRE) could also impact spin pumping experiment results. Permalloy is the most widely used magnet for the spin pumping experiments. We also conducted a control spin pumping experiment with voltage on Py/Cu, which shows negligible response, ruling out SRE contribution.

The NiFe/AlOx/KTaO$_3$(111) sample shows a clear spin-to-charge conversion response when spin pumped using the magnetic NiFe overlayer. The spin wave is converted into a charge current through the inverse Edelstein effect in the KTaO$_3$(111) 2DEG due to the Rashba splitting of the 5$d$ orbital (Figure 1(a)). The spin pump voltage shows a linear response to the power and increases with decreasing temperature. The spin-to-charge conversion efficiency which is calculated using the measured damping factor gives a value of ~0.6 nm at 10 K. The $\lambda_{IREE} = \alpha_R \tau/\hbar$, where $\alpha_R$ is Rashba parameter. Using the mean free time of charge carriers at low temperature (~0.02 ps at 3K), $\alpha_R = 2\ meV$ Å. This value matches the calculated Rashba spin splitting along [11$\bar{2}$] direction[6].

Next, we measured the angle dependence of spin-to-charge conversion response. Figure 4(a) shows the geometry of the experiment. The in-plane magnetic field was rotated in a vector magnet and the experiment was repeated at various angles. Figure 4(b) and (c) show the spin pumping response as a function of in-plane field angle at 10 K and 70 K, respectively. In a simple heavy-metal/ferromagnet bilayer system (e.g., Py/Pt) with no broken lateral symmetries a $\cos(\varphi)$ behavior is expected[34,35]. Here, there are clear deviations from this description. This could be due to the deviation of spin texture from ideal Rashba picture[18]. Combining the trivial results with a $\cos(3\varphi)$ term could describe the angular dependence of experimental results. The KTaO$_3$ (111) surface has C$_{3v}$ group symmetry. Figure 1 exhibits the Fermi surface calculation for 50 meV and 100 meV chemical potentials, reflecting the C$_{3v}$ group symmetry which may explain cos(3$\varphi$) behavior.

First, the spin-to-charge conversion response has a trigonal symmetry (cos(3$\varphi$)) which matches the symmetry of the Fermi states, Figure 1(a). The spin to charge conversion current can be described as: j$_k$ = $\gamma_{kc}$ $\delta S_c$, where $\gamma_{kl}$ is a second rank pseudo tensor. Since KTaO$_3$ has a crystal symmetry belonging to the C$_{3v}$ group, the only nonzero elements of this tensor are $\gamma_{xy}$ = $-\gamma_{yx}$ =: $\gamma$. Thus, j$_x$ = $\gamma$ $\delta S_y$. For this reason, the response is maximal if the magnetic field is perpendicular to the transport direction but vanishes when magnetic field and transport direction are parallel. This explains why the voltage across the junction has a maximum for $\phi \approx 0°$ and vanishes for $\phi \approx 90°$. However, because the relation is linear, the spin to charge conversion effect cannot be responsible for the higher harmonics that are present the signal in the dependence of voltage on magnetic field. Thus, there must be a non-linearity in the relation between the induced spin accumulation and the magnetic field, that leads in Eq. (29). There are a few ways in which such non-linearities can appear, (i) the applied magnetic field and the induced spin are not parallel, (ii) the magnitude of the induced spin accumulation might depend on the direction of the magnetic field and (iii) both the spin angular momentum (SAM) and orbital angular momentum (OAM) contribute significantly, and finally (iv) anisotropic magnetoresistance (AMR) signal is not negligible.

Per mechanism (i) the calculated Fermi surface and band structure demonstrate the anisotropy of the spin texture in a KTaO$_3$(111) 2DEGs, including substantial deviation from ideal Rashba spin-momentum locking with momentum-dependent out-of-plane spin canting[6,18]. Furthermore, we notice slight deviation in which the largest response does not occur at zero angle. This could be a signature of the deviation of spin texture from ideal Rashba picture. This mechanism, however, provides small corrections and does introduce extra minima, as this would require a non-monotonic dependence of the angle of the induced spin with the y-axis on the angle of the magnetic field with the y-axis.

Mechanism (ii) requires that the magnitude of the induced spins depend on the direction of the magnetic field. For example, a function that gives similar results to the fits is |$\delta$S| = 1/2(1 + cos(3$\phi$)) cos($\phi$)f(|B|), where f(|B|) is a nonlinear function of the magnitude of the magnetic field. This gives $\delta S_y(\phi)$ = 1/2 (1 + cos 3 $\phi$) cos($\phi$)f(|B|). This function, however, implies that $\delta S_y(\phi) \neq -\delta S_y(\phi + \pi)$, while by definition V$_{IEE}$ =V$_{sym}$(+H) − V$_{sym}$(−H) and as a result $\delta S_y(\phi)$ = $-\delta S_y(\phi + \pi)$, as illustrated in Figure 3(a).

Mechanism (iii) suggests SAM and OAM both play a significant role. Fourfold symmetry of the spin to charge current conversion in a SrTiO$_3$/LaAlO$_3$ (001) 2DEGs[36,37]. This anisotropy is associated with combination of SAM and OAM contributions. El-Hamdi et al., argue that the nature of tunnelling barrier, i.e., epitaxial LaAlO$_3$, is critical in allowing the observation of electronic symmetry[36]. Here, however, despite having an amorphous AlOx barrier, we observe a symmetry that mimics the Fermi state. Finally, a trivial explanation of the observed signal could rely on strong anisotropic magnetoresistance (AMR) contribution. Angle dependence of spin pumping in Py/Pt were described using AMR angle dependence which follows a $Sin(2\phi) \times Sin(\phi)$ behavior[38].

Finally, we discuss the relatively weak spin-to-charge conversion response in $KTaO_3$ 2DEGs compared to $SrTiO_3$ despite the large SOC in tantalum $5d$ driven conduction band[3]. First, we stress that the large SOC in tantalum $5d$ driven states, compared to titanium $3d$, does not directly translate into stronger Rashba SOC. Inefficient spin-to-charge conversion, compared to $SrTiO_3$, was previously reported in $KTaO_3$ (001) 2DEGs[16]. Here, orbital symmetry could lead to compensation effects, suppressing the spin-charge interconversion efficiency. The compensation could be due to opposite sign of SAM and OAM contributions[36] and/or opposite sign of contributions from different orbitals ($d_{xy}$, $d_{xz}$ and $d_{yz}$)[39]. Furthermore, it was demonstrated that the $J=3/2$ states in $KTaO_3$ near gamma point have anomalous quasiparticles with vanishing magnetic moment ($\hat{m} = -\mu_o(\hat{l} + 2\hat{s}) \approx 0$)[7], which could also lead to inefficient spin-to-charge conversion.

Applying the Klemm-Luther-Beasley relation[40] for $KTaO_3$ (111) superconducting transport results reported earlier[18] exhibits that spin-orbit time is much longer than the electronic transport time from Hall experiment ($\tau_{SO}/\tau_{tr} \sim 10$), suggesting that the relaxation time is dominated by the electronic transport and not the spin-orbit scattering events. As a result, we expect that increasing the interfacial carrier mobility, which here is currently below the bulk values (>10,000 cm$^2$/V.s[25,41]), will enhance the spin-to-charge conversion efficiency.

In summary, our results demonstrate efficient spin-to-charge conversion. The conversion efficiency is not scaled with spin-orbit coupling compared to similar systems, e.g., $SrTiO_3$. Here, the angle dependence of the inverse Rashba-Edelstein response mimics the symmetry of the Fermi states. Future experiments using microfabricated devices along various crystallographic directions could further elucidate the nature of Rashba-Edelstein effect's dependence on calculated OAM and SAM components. Results point to opportunities to harness Rashba-Edelstein effect to probe electronic structure and spin texture in 2D electron systems with enhanced spin-orbit coupling.

Ohio State team was supported by the U.S. National Science Foundation (NSF) under Grant DMR-2408890. D.S. acknowledges the financial support by the U.S. National Science Foundation (NSF) under Grant DMR-2143642. Authors acknowledge Fermi surface spin texture calculations (Figure 1a) by M.N. Gastiasoro. KA acknowledges discussions with J. Lorenzana, M.N. Gastiasoro, Ilya Tokatly, and Sebastian Bergeret.

The data that support the findings of this study are available in the article. Raw data can be obtained from the corresponding author upon request.

The authors declare that they have no conflict of interest.

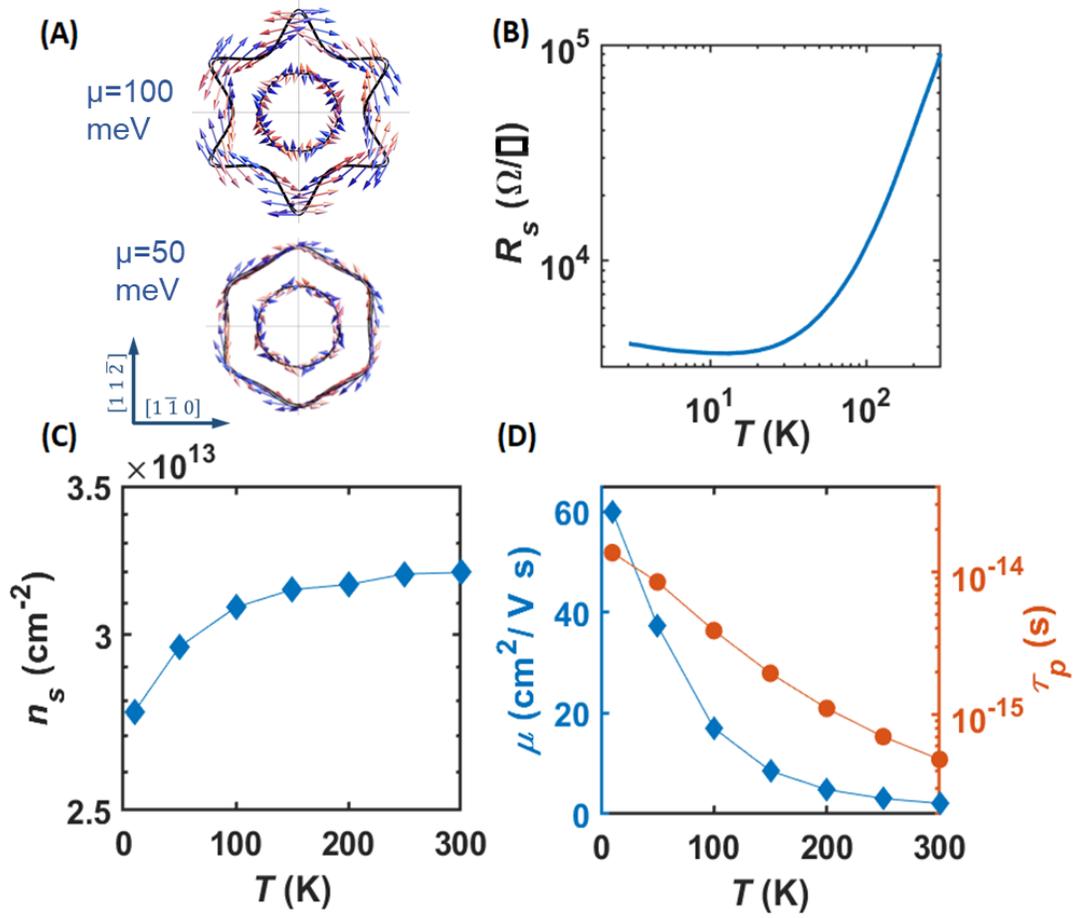

**Figure 1. Electronic structure and transport at the KTaO$_3$(111) interfaces.** (A) Fermi surface and Rashba spin texture of the 2DEG at the KTaO$_3$(111) interface for chemical potentials µ = 50 meV (bottom) and µ = 100 meV (top). The calculation details were described elsewhere[7]. (B) Sheet resistance of KTaO$_3$ (111) 2DEG with temperature. (C) Hall sheet carrier density with temperature. (D) Carrier mobility and mean free time with temperature. Mean free time was calculated using Hall mobility and assuming effective mass of 0.52[12].

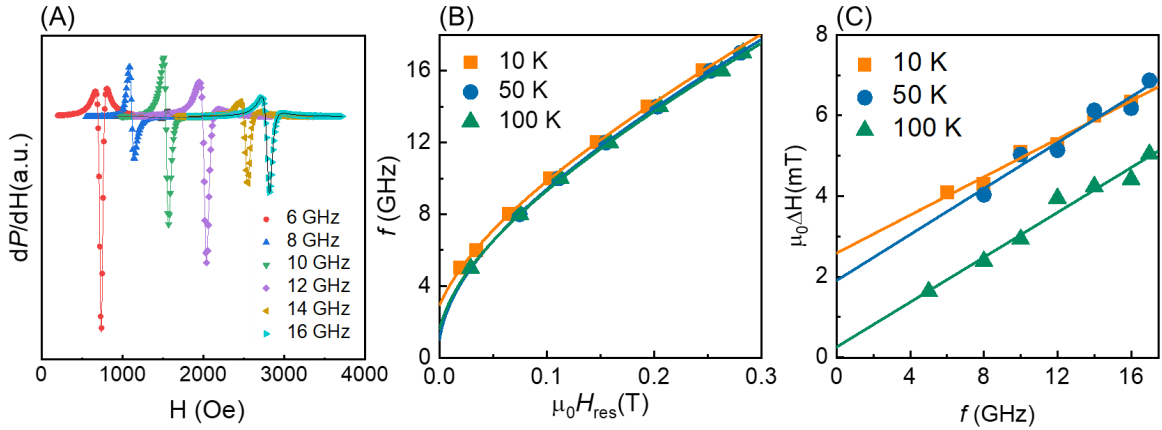

**Figure 2. FMR response of NiFe/AlOx/KTaO₃(111) heterostructure.** (A) The representative FMR spectra at 10 K. (B) The typical relation between resonance field $\mu_0 H_{res}$ and frequency, which can be described by the Kittel equation (solid curves). (C) The obtained resonance linewidth $\mu_0 \Delta H$ at respective f. The slope represents the effective damping factor.

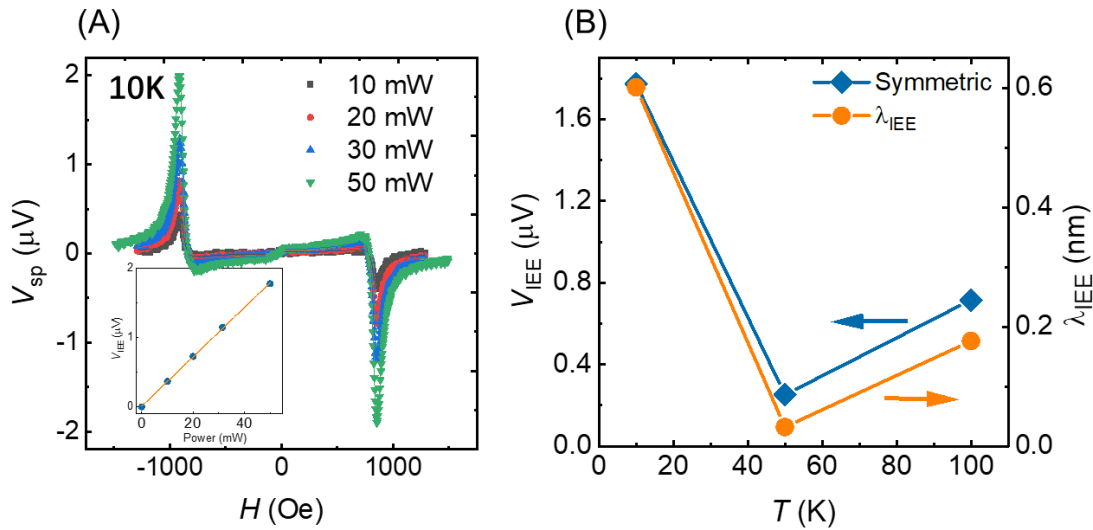

**Figure 3. Spin pumping response of NiFe/AlOx/KTaO₃(111) at 10 K.** (A) Microwave power-dependent spin pumping response as a function of magnetic field. The inset shows the linear relation between mcirowave power and IEE response. (B) The IEE response voltage and the spin-charge conversion efficiency $\lambda_{IEE}$ at different temperatures.

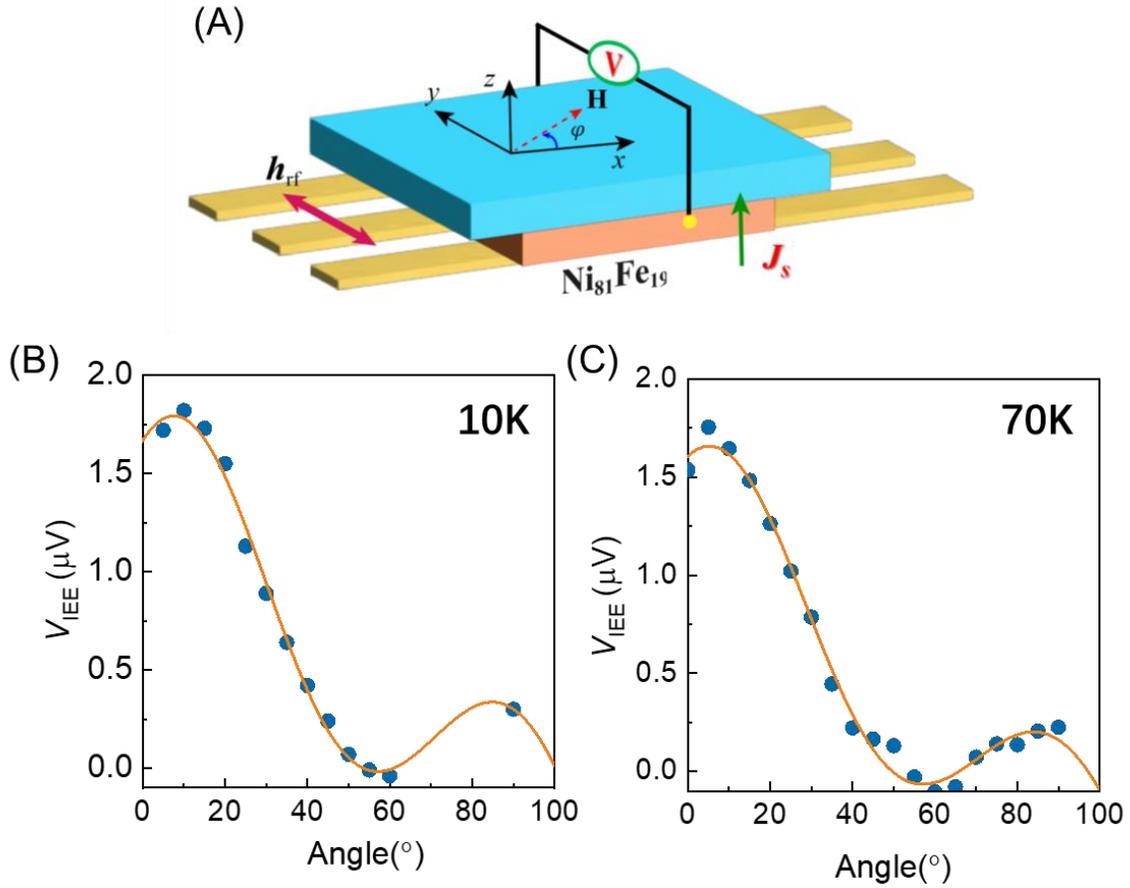

**Figure 4. Angle dependence of spin pumping response in NiFe/AlOx/KTaO₃(111) heterostructure.** (A) The schematic diagram of in-plane angular dependence of spin pumping measurement. Angle dependence of the spin pumping response $V_{sp}$ at (C) 10 K and (D) 70 K. $A cos(\varphi) cos(3\varphi) + C sin(\varphi) cos(3\varphi)$. $Cos(\varphi)$ represents the traditional IEE response from the perpendicular spin-momentum locking; $Cos(3\varphi)$ represents the trigonal symmetry; $Sin(\varphi)$ term accounts for the deviation from Rashba spin-momentum locking, including a reported momentum-dependent out-of-plane spin canting[6,18]. The $\left|\frac{A}{C}\right|$ ratio is 1.9 and 2.8 at 10 K and 70 K, respectively.